\documentclass[10pt,twocolumn,floatfix,pra]{revtex4-1}
\usepackage{amsmath,amssymb,amsthm,mathrsfs,amsfonts,dsfont,amstext} 
\usepackage{textcomp,pbox}
\usepackage[export]{adjustbox}
\usepackage{bm}
\usepackage{dcolumn,booktabs,url}
\usepackage[scaled]{helvet}
\usepackage{sansmath,gensymb}
\usepackage{tikz,graphicx,transparent,color}
\usepackage{multirow}
\usepackage{braket}

\usepackage[colorlinks=true]{hyperref}
\usepackage{graphicx}

\graphicspath{{./figures/}}

\newcommand{\yso}{Y$_2$SiO$_5$}

\newcommand{\ndiso}[0]{$^{145}$Nd$^{3+}$:Y$_2$SiO$_5$}
\newcommand{\transition}{$^4$I$_{9/2}(0) \longleftrightarrow ^4$F$_{3/2}(0)$ }

\begin{document}

\newcommand{\TitleName}{Efficient optical pumping using hyperfine levels in \ndiso{} and its application to optical storage}
\title{\TitleName}

\author{Emmanuel Zambrini Cruzeiro$^{1}$}
\author{Alexey Tiranov$^{1}$}
\author{Jonathan Lavoie$^{1}$}
\altaffiliation{Present address: Department of Physics and Oregon Center for Optical Molecular \& Quantum Science, University of Oregon, Eugene, OR 97403, USA}
\author{Alban Ferrier$^{2,3}$}
\author{Philippe Goldner$^{2}$}
\author{Nicolas Gisin$^{1}$}
\author{Mikael Afzelius$^{1}$}
\email{mikael.afzelius@unige.ch}

\affiliation{$^{1}$Groupe de Physique Appliqu\'ee, Universit\'e de Gen\`eve, CH-1211 Gen\`eve, Switzerland}
\affiliation{$^{2}$Chimie ParisTech, PSL University, CNRS, Institut de Recherche de Chimie Paris, 75005 Paris, France }
\affiliation{$^{3}$Sorbonne Universit\'{e}, Facult\'{e} des Sciences et Ing\'{e}nierie, UFR 933, Paris, France
}

\begin{abstract}
Efficient optical pumping is an important tool for state initialization in quantum technologies, such as optical quantum memories. In crystals doped with Kramers rare-earth ions, such as erbium and neodymium, efficient optical pumping is challenging due to the relatively short population lifetimes of the electronic Zeeman levels, of the order of 100 ms at around 4 K. In this article we show that optical pumping of the hyperfine levels in isotopically enriched \ndiso{} crystals is more efficient, owing to the longer population relaxation times of hyperfine levels. By optically cycling the population many times through the excited state a nuclear-spin flip can be forced in the ground-state hyperfine manifold, in which case the population is trapped for several seconds before relaxing back to the pumped hyperfine level. To demonstrate the effectiveness of this approach in applications we perform an atomic frequency comb memory experiment with 33\% storage efficiency in \ndiso{}, which is on a par with results obtained in non-Kramers ions, e.g. europium and praseodymium, where optical pumping is generally efficient due to the quenched electronic spin. {Efficient optical pumping in neodymium-doped crystals is also of interest for spectral filtering in biomedical imaging, as neodymium has an absorption wavelength compatible with tissue imaging. In addition to these applications, our study is of interest for understanding spin dynamics in Kramers ions with nuclear spin.}
\end{abstract}

\maketitle

\section{Introduction}

In rare-earth-ion doped crystals, one often uses frequency-resolved optical pumping (i.e. spectral hole burning) as a necessary preparation step in various applications, such as optical quantum memories~\cite{Tittel2010b,Afzelius2009a,Hedges2010,Jobez2016}, spectrum radio-frequency analysers~\cite{Babbitt2007,Linget2015}, and narrow spectral filters for biological imaging~\cite{Zhang2012,McAuslan2012a}. {For the latter application, there is a special interest in materials such as YSO doped with neodymium and thulium \cite{Walther2017}}. Common to all these techniques is the need for creating narrow and deep spectral holes in the inhomogeneously broadened optical transition. The most efficient way for realizing deep spectral holes is to optically pump ions into a long-lived ground state, which could be electronic spin or nuclear spin states. Generally it is observed that the efficiency of the optical pumping, i.e. the depth of the spectral hole, strongly depends on the ground-state population lifetime with respect to the radiative lifetime~\cite{Lauritzen2008,Afzelius2010b,Saglamyurek2015,Zhong2017}.

In non-Kramers ions, such as europium, praseodymium or thulium, the ground state is an electronic singlet in low-symmetry doping sites, such that only nuclear interactions (eg. Zeeman and quadrupole) exist in the ground state~\cite{MacfarlaneShelby1987,Macfarlane2002}. As a result both spin-spin interaction and spin-lattice relaxation rates are low, with lifetimes of many seconds or hours~\cite{Macfarlane2002}, and efficient optical pumping can be achieved in a wide range of experimental parameters (in terms of doping concentration, temperature and magnetic field). 

In Kramers ions such as erbium, neodymium or ytterbium, the ground state is an electronic Zeeman doublet $S=1/2$ in low-symmetry doping sites, with a magnetic moment in the range of 1--15 $\mu_B$, where $\mu_B=14$~GHz/T is the Bohr magneton. The large moment results in strong dipole-dipole interactions between Kramers ions and faster spin-lattice relaxation rates. In terms of applications, however, Kramers ions are interesting as the large Zeeman and hyperfine splittings allow large bandwidth quantum memories~\cite{Saglamyurek2011,Clausen2011,Zhong2017} and could be interfaced with superconducting qubits working in the 1-10~GHz regime~\cite{Staudt2012,Probst2013}.

Recently we presented a detailed study of the spectral hole lifetime in naturally doped Nd$^{3+}$:Y$_2$SiO$_5$~\cite{ZambriniCruzeiro2017a}, as a function of magnetic field strength and direction, temperature (between 3 and 5.5 K) and Nd$^{3+}$ doping concentration. It was found that the spectral hole dynamics {were} dominated by the population relaxation between the electronic Zeeman states $m_S = \pm 1/2$ of even Nd$^{3+}$ isotopes having zero nuclear spin $I=0$. For doping concentrations required for practical applications (a few tens of ppm) the longest measured lifetime was around 160 ms. This relatively short spectral hole lifetime results in residual absorption backgrounds of about 6--7\% of the peak absorption, which has been the main factor limiting the efficiency in several quantum storage experiments we have carried out in naturally doped Nd$^{3+}$:Y$_2$SiO$_5$~\cite{Clausen2011, Usmani2012, Froewis2017}. Such background absorptions would not allow very efficient (50\% or above) quantum memories based on cavity enhancement~\cite{Moiseev2010a,Afzelius2010a}, as demonstrated in praseodymium and europium doped materials~\cite{Sabooni2013,Jobez2014}. Also, there is an interest in using the 883 nm resonance in Nd$^{3+}$:Y$_2$SiO$_5$ crystal for biological imaging~\cite{Walther2017}, but spectral filtering with high dynamic range requires a lower residual absorption. Hence, further application of Nd$^{3+}$-doped materials in these research areas would greatly benefit from longer spectral hole lifetimes.

In naturally doped Er$^{3+}$:Y$_2$SiO$_5$ similar spectral hole lifetimes of up to 130~ms have been achieved using isotopes with $I=0$, at very low magnetic fields and around 3~K~\cite{Hastings-Simon2008a}. Longer lifetimes can be achieved both in Er$^{3+}$:Y$_2$SiO$_5$ crystals~\cite{Probst2013} and Er$^{3+}$-doped fibers~\cite{Saglamyurek2015}, but at sub-K temperatures. At temperatures above 1~K, Ran\v{c}i\'{c} and co-workers recently demonstrated long spectral hole lifetimes of up to 60~seconds in a isotopically enriched $^{167}$Er$^{3+}$:Y$_2$SiO$_5$ crystal~\cite{Rancic2017}, where $^{167}$Er has a nuclear spin $I=7/2$. In their approach spin-lattice relaxation between the electronic Zeeman states $m_S=\pm1/2$ is suppressed by applying a high magnetic field of 3~T or more, in which case the hole lifetime is dominated by relaxation between nuclear states $m_I$ in the lowest electronic Zeeman state. A similar suppression of the electronic spin-lattice relaxation in Nd$^{3+}$-doped samples would require much higher magnetic fields, or much lower temperatures, due to four times lower magnetic moment of Nd$^{3+}$ ions. 

As an alternative approach we explore spectral hole lifetimes in moderate magnetic fields and a temperature of about 3~K in an isotopically enriched $^{145}$Nd$^{3+}$:Y$_2$SiO$_5$ crystal, detailed in Sec.~\ref{sec:exp_details}, where $^{145}$Nd also has a nuclear spin $I$=7/2. In Sec.~\ref{sec:SHBmeasurements} we show that the spectral hole lifetime depends strongly on the duration of the hole burning pulse. By burning for durations of around one second, the spectral hole is dominated by a slow decay process with a lifetime that can reach 4~seconds.  In Sec.~\ref{sec:relax_model} we develop a simple model to explain our data, where the basic idea is that optical pumping for long durations increases the probability of flipping the nuclear spin projection $m_I$ in the ground state. Once the nuclear spin has flipped through the optical excitation, the spin flip-flop and spin-lattice relaxation rates are reduced, as these rates depend on nuclear spin mixing induced by the non-secular part of the hyperfine interaction $S \cdot A \cdot I$. In Sec.~\ref{sec:AFC_storage_application} it is shown that the efficiency of the optical pumping is enhanced as a result of the longer hole lifetimes, with an absorption background of 1-2\% of the peak absorption. By applying these results to an optical storage experiment based on an atomic frequency comb memory, we reach storage efficiencies of up to 33\%, which is mainly limited by the peak absorption rather than the residual background absorption. 

\section{Experimental details}
\label{sec:exp_details}

\begin{figure}[h]
\begin{center}
\includegraphics[width=1\linewidth,trim= 0.3cm 0.2cm 0.5cm 0.5cm, clip] {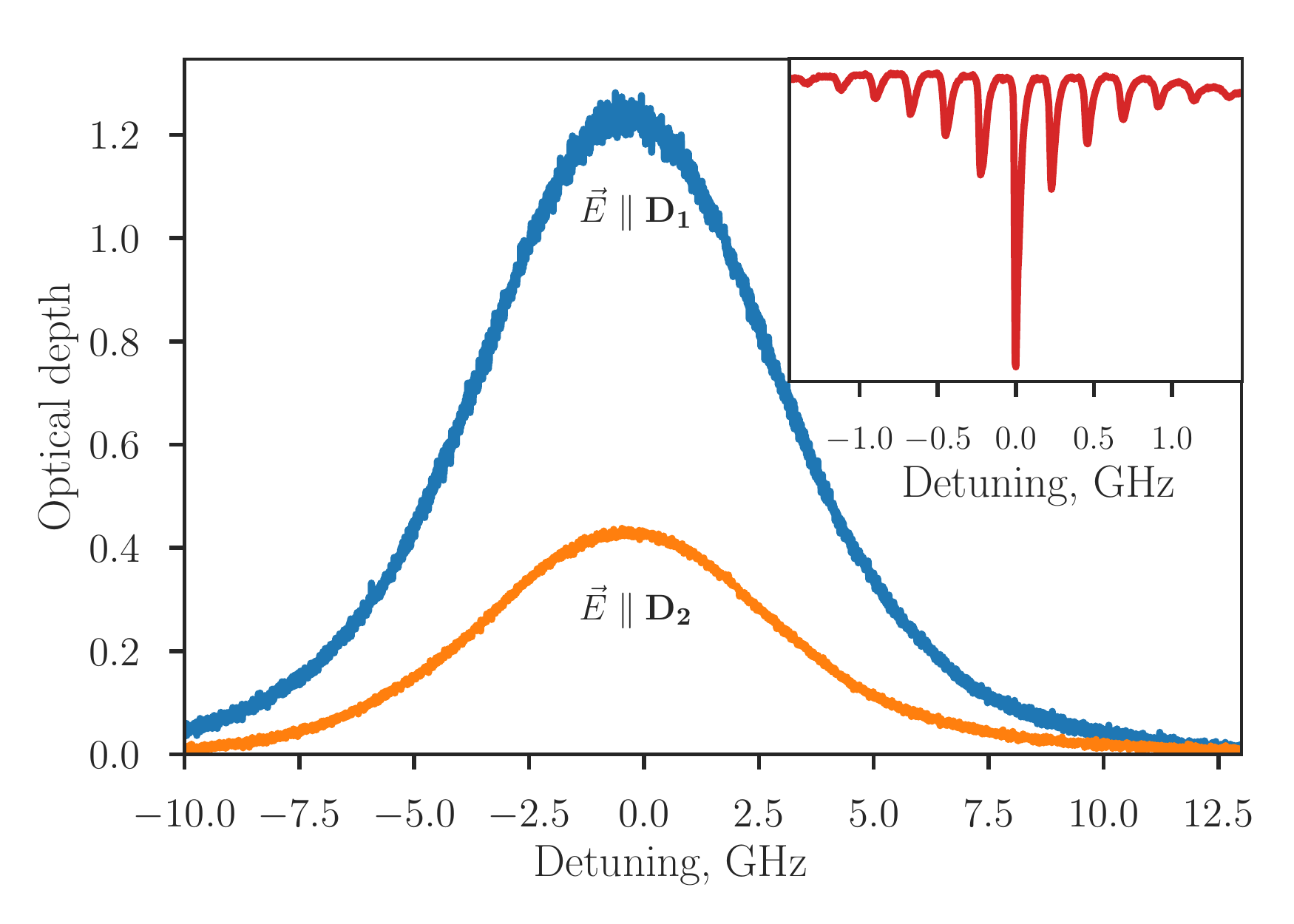}
\caption{The optical inhomogeneous absorption profile on the \transition{} transition, at a temperature of 3~K and with no applied magnetic field. The measured peak optical depth is 1.26 (0.44) for the light polarization along the $D_1$ ($D_2$) axis, resulting in an absorption coefficient $\alpha$ = 1.05~cm$^{-1}$ ($\alpha$ = 0.37~cm$^{-1}$). The inhomogeneous lineshape is Gaussian with a FWHM linewidth of 7.7 GHz. In the inset we show the holeburning spectrum showing the central hole and the associated side holes due to the hyperfine splitting in the optically excited state $^4$F$_{3/2}$.}
\label{fig:abs_spectrum}
\end{center}
\end{figure}

Yttrium orthosilicate \yso{} crystal belongs to the crystallographic group $C_{2h}^6$. It is known as a host material that provides excellent coherence properties due to its low nuclear spin density. Neodymium ions can substitute yttrium ions in two crystallographic sites, both of low $C_1$ symmetry~\cite{Maksimov1970}. We grew the crystal by the Czochralski
method using the parameters given in \cite{Ferrier2016}. It was doped with 0.001 at.\% concentration of $^{145}$Nd$^{3+}$ ions, with an isotopic purity of about 90\%. We note that crystals from the same boule have been used in electron spin measurements presented in Refs~\cite{MaierFlaig2013,Wolfowicz2015}.  Our sample was cut with faces perpendicular to the $b$, $D_1$, and $D_2$ optical extinction axes~\cite{Li1992}, where light was propagating along the $b$ axis (of length 12 mm). The states of interest in this work are the ground $^4$I$_{9/2}$ and excited $^4$F$_{3/2}$ states of site 1. In the $C_1$ site symmetry, the ground and excited states split into 5 and 2 Kramers doublets, respectively. All measurements presented here were carried out on the transition between the Kramers doublets of lowest energy, at a wavelength of 883.0~nm (in vacuum) for site~1~\cite{Beach1990}.

The optical measurements were performed with an external cavity diode laser {with $\approx$1~MHz linewidth and drift within the tens of MHz per hour}. Large bandwidth frequency scans (in the GHz range) were performed by scanning a piezo element in the cavity. An acousto-optical modulator (AOM) in double pass configuration provided fine control of the amplitude and frequency of the light (120 MHz bandwidth). The AOM was used to create the optical pulses for the spectral holeburning measurements, as well as for preparing the atomic frequency comb in the light storage experiment.

In Figure \ref{fig:abs_spectrum} we show the optical absorption spectrum recorded at 3~K and zero applied magnetic field. The strongest absorption is obtained with the linear light polarization along the $D_1$ axis. The absorption profile follows a smooth distribution, without any visible substructure due to the hyperfine splittings of the ground and excited states. To reveal transitions hidden within a large inhomogeneous absorption profile one can employ spectral hole burning (SHB) techniques~\cite{MacfarlaneShelby1987}. A typical SHB measurement {performed at the center of the absorption line} consists of a burn pulse at a fixed frequency and a time-delayed probe pulse whose frequency is scanned around the burn pulse frequency. The spectral hole appears at the burn frequency, while side holes appear at frequencies given by any splittings in the optically excited state~\cite{MacfarlaneShelby1987}. In the inset of Figure~\ref{fig:abs_spectrum} we show the SHB spectrum in \ndiso{} with zero applied magnetic field. A set of periodic side holes can indeed be observed, due to the hyperfine levels in the excited state. Unfortunately the large number of expected side holes due to the $(2S+1)\times(2I+1)=16$ hyperfine levels does not allow any quantitative analysis of the hole structure. However, the exact SHB spectrum is not of interest here, as we will principally investigate the time-resolved dynamics of the central hole in the SHB spectrum.

\section{Spectral hole decay measurements}
\label{sec:SHBmeasurements}

The SHB mechanics of interest here is due to trapping of population in ground state hyperfine levels that are not excited by the burn pulse in the SHB sequence. In this way the observed hole decay curve contains information about different relaxation rates within the ground state hyperfine levels. This regime is achieved by optically burning the hole for a duration much longer than the excited state lifetime, which for $^4$F$_{3/2}$ is 225 $\mu$s~\cite{Beach1990}. In the experiments presented here the burn pulse duration varied between 10 ms and 950 ms. To measure the decay of the hole, we varied the delay $t_d$ between the burn and probe pulses and recorded the area of the central spectral hole for each delay. By using the hole area the measurement is independent of spectral diffusion {\cite{Bottger2009}}. A range of SHB decay measurements were made with different burn pulse durations and applied magnetic fields. For all measurements, however, the magnetic field orientation was along the $D_1$ axis and the temperature was~3~K.

\begin{figure}[h]
\begin{center}
\includegraphics[width=1\linewidth] {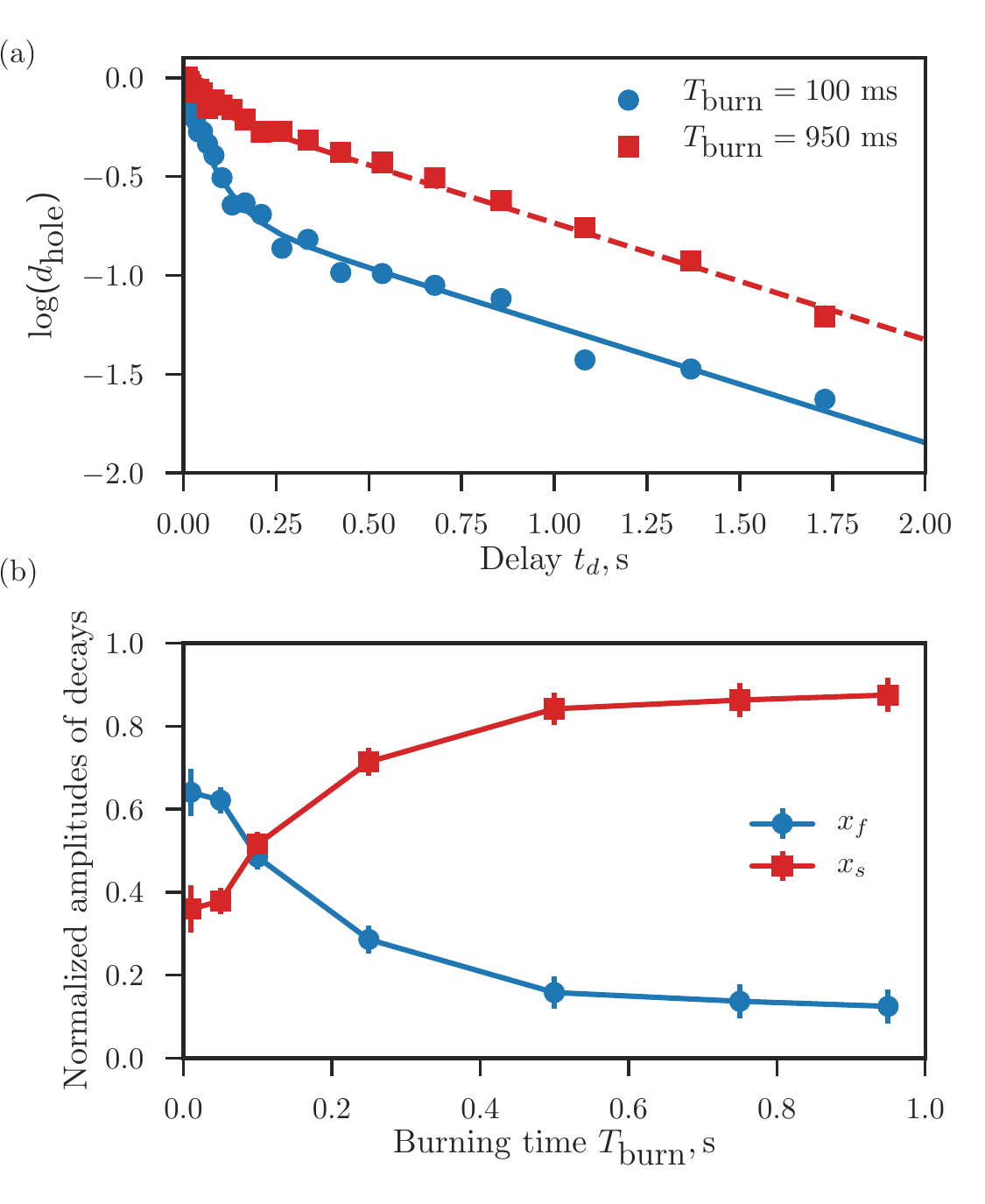}
\caption{(a) The spectral hole amplitude as a function of the delay between the burn and probe pulses, for two burn durations of $T_{burn}$ = 100 ms (circles) and 950 ms (squares), respectively. The vertical axis shows the logarithm of the optical depth in the spectral hole. Each curve has been shifted to zero at the shortest delay. The solid lines are fitted curves using the simple model described in the text. (b) Relative amplitudes of the fast (circles) and slow (squares) decay constants, as a function of the burn duration $T_{burn}$ (see text for details).}
\label{fig:decaycurves_amps}
\end{center}
\end{figure}

All SHB decay measurements featured the same overall trend of a fast decay in the beginning, followed by a much slower decay of the hole. A striking observation was that the relative amplitudes of the fast and slow decays depended on the duration of the burn pulse $T_{burn}$, while the time constants of the fast and slow decays were independent of the burn pulse duration. In Figure \ref{fig:decaycurves_amps}(a) we show two typical SHB decay curves for a shorter (100 ms) and longer (950 ms) burn pulse, at a magnetic field of 900 mT. In this particular case the fast decay constant was 75 ms, while the slow decay constant was 1.72 s. It is also clearly seen that the fast decay dominates the decay for a short burn pulse, while the slow decay dominates the decay for the long burn pulse.

To investigate the dependence of the relative amplitudes of the fast and slow decays on the burn duration, we recorded and analysed an entire set of decay curves for $T_{burn}$ in the range of 10 to 950 ms. Each set was analysed using only two time constants that were common to all curves, while the relative amplitudes were fitted individually. Specifically, the decay of the spectral hole depth $d_{hole}$ was modelled as $d_{hole}(t_d) = d_f \exp(-t_d/T_f) + d_s \exp(-t_d/T_s)$. Here $d_f$ and $d_s$ are the amplitudes, and $T_f$ and $T_s$ the time constants, of the fast and slow decays, respectively. The parameters $d_f$ and $d_s$ were fitted for each burn duration, while $T_f$ and $T_s$ were fitted globally for the entire set of burn durations for a given magnetic field strength. We emphasize that all sets of decay curves could be analysed with only two time constants, for all magnetic fields.

In Figure  \ref{fig:decaycurves_amps}(b) we show the relative amplitudes of the fast and slow decays, which we define as $x_f=d_f/(d_f+d_s)$ and $x_s=d_s/(d_f+d_s)$, respectively, as a function of the burn duration for a field of 900 mT. By pumping for sufficiently long, one is able to force the hole to decay predominantly through a slower relaxation mechanism. The crossing point in this particular case is for a burn duration of about 90 ms. In the following section we will present a simple relaxation model to explain the observed SHB curves.

\section{Hyperfine relaxation model}
\label{sec:relax_model}

\begin{figure}[htb]
\begin{center}
\includegraphics[width=1\linewidth] {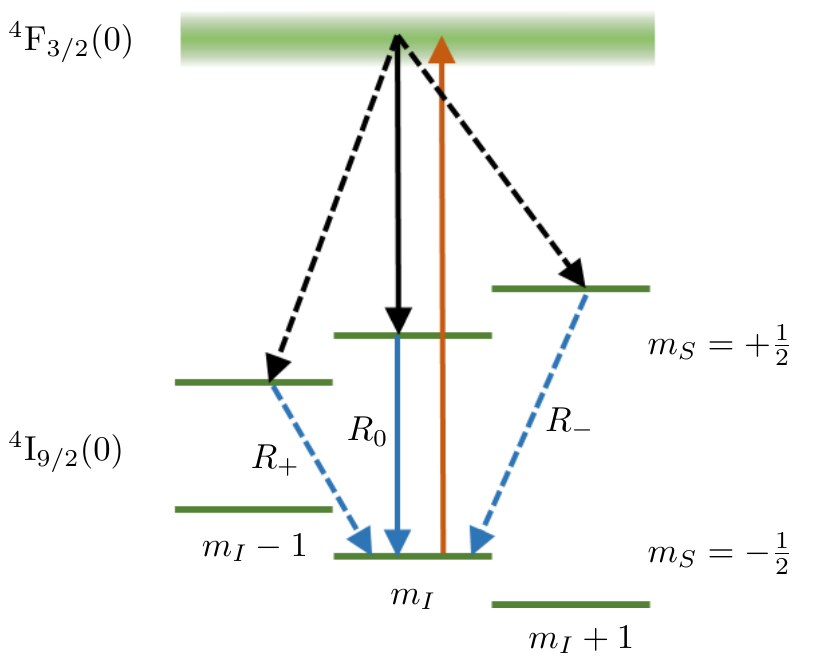}
\caption{Simplified energy level diagram of $^{145}$Nd$^{3+}$ ~for understanding the simple relaxation model presented in Sec. \ref{sec:relax_model}. The optical excitation is shown as the red, solid line. The most probable decay path from the excited state is shown as a black solid line, while the less probable paths that change nuclear spin projection are shown as black dashed lines. The fast and slow spin relaxation rates are shown as blue solid and dashed lines, respectively. Three different spin relaxation rates $R_{0}$, $R_{+}$ and $R_{-}$ are defined. For most of the magnetic fields studied in this article, the ground state splits into two electronic branches ($m_S = \pm 1/2$), with eight possible nuclear spin projections for each branch ($m_I = -7/2,..,7/2$). For simplicity only three nuclear spin projections are shown. The hyperfine structure of the excited state is not discussed here, hence it is shown as a fuzzy region.}
\label{fig:model}
\end{center}
\end{figure}

The spectral hole is created by a re-distribution of population among the hyperfine levels in the ground state of $^{145}$Nd$^{3+}$, with respect to the thermal distribution, due to the cycling of population through the optically excited state. {At most magnetic fields used here, the ground state Zeeman splitting between the $m_S=\pm 1/2$ branches is larger than the optical broadening, such that the corresponding states can be resolved. For those fields we consistently excited the $m_S=- 1/2$ ground state, as shown in Fig. \ref{fig:model}. For weaker fields, both ground states were excited simultaneously. We remark, however, that in our recent study of Nd$^{3+}$ ions having no nuclear spin~\cite{ZambriniCruzeiro2017a}, the spectral hole decay did not depend on the initial state $m_S=\pm 1/2$.} The split of the hyperfine states within a given $m_S$ branch, on the other hand, is not resolved optically, as exemplified by the smooth zero-field absorption spectrum shown in Fig. \ref{fig:abs_spectrum}. The spectral hole is thus an average of optical pumping of all possible initial states~$m_I$. In addition, in this regime of magnetic fields the hyperfine interaction is a weak perturbation to the Zeeman interaction. Then the separable states $\ket{m_S}\ket{m_I}$ describe, to a good approximation, the hyperfine levels in the electronic ground state (see Fig. \ref{fig:model}). However, as we will see, a small mixing of states $\ket{m_S}\ket{m_I}$ due to the hyperfine interaction is essential in order to understand the different spin relaxation rates.

As population is optically cycled through the excited state, ions can {optically} decay back to different hyperfine {ground state} levels. If the ion {optically decays into} the opposite $m_S = +1/2$ {ground} state having the same $m_I$, then it can {relax} through the process that does not involve any change in the nuclear spin projection $\Delta m_I = 0$. We denote this {ground-state} relaxation rate $R_0$, as shown in Fig.~\ref{fig:model}. {For a magnetic field of 900 mT for which the results} are shown in Fig. \ref{fig:decaycurves_amps}, the dominant relaxation process is the direct spin lattice relaxation (SLR)~\cite{ZambriniCruzeiro2017a}. For the SLR process this relaxation path is expected to have the highest rate (shortest lifetime) ~\cite{Larson1966a}, due to the low spin mixing. The hyperfine Hamiltonian $S \cdot A \cdot I$, does however open up other relaxation paths, owing to non-secular mixing terms such as $S_{-}I_{+}$ and $S_{-}I_{-}$. We denote these relaxation rates by $R_{+}$ and $R_{-}$, which are expected to be significantly slower than $R_0$~\cite{Larson1966a}. {In the simpler case of an axially symmetric hyperfine Hamiltonian with the field along one of the principal axes, then $R_+=R_-=\left(\frac{A}{\Delta E_g}\right)^2R_0$ {\cite{Abragam1970_page78}} where $A$ is the hyperfine component perpendicular to the magnetic field and $\Delta E_g$ is the electronic ground state Zeeman splitting}. Other, pure nuclear relaxation paths, for which $\Delta m_S = 0$, are assumed to be even slower~\cite{Rancic2017} and are not considered here.

Using this simple model we can make sense of the spectral hole decay curves shown in Fig. \ref{fig:decaycurves_amps}, by additionally assuming that the optical decay preferably preserves the nuclear spin projection $m_I$, as shown by the solid line in Fig. \ref{fig:model}. A shorter burn pulse would then preferentially optically pump ions into the opposite $m_S$ state with identical $m_I$ projection, yielding a fast hole relaxation given by $R_0$. By increasing the burn duration we can increase the chance of optically decaying into neighbouring $m_I$ states, as shown by the dashed lines in Fig. \ref{fig:model}, which would yield slower hole decay rates of the order of $R_{+}$ and/or $R_{-}$. In principle, we should then observe two additional rates, while experimentally one slow decay constant is observed. However, in theory the rates $R_{+}$ and $R_{-}$ also have $m_I$ dependence, which is also different for the two rates, as discussed by Larson and Jeffries~\cite{Larson1966a}. In our case we believe that either we measure an effective average rate, which would correspond to the slow decay constant $T_s$, or one of the rates dominate over the other for the particular field orientation used in this experiment. Further experimental measurements and a detailed theoretical modelling would be necessary to decide upon this question. Yet, the overall behaviour is consistent with this simple model.

To better understand the relaxation mechanisms governing the fast and slow decays, we measured the hole decay curve for fields in the range of 10 mT to 1.6 Tesla. But before discussing these results, we briefly recall the main findings of Ref.~\cite{ZambriniCruzeiro2017a} where we studied the decay of spectral holes due to relaxation between $m_S = \pm 1/2$ states for ions with no nuclear spin $I=0$. At low fields (around 0.5 Tesla) the decay was given by spin-spin relaxation, with a linear increase of the spectral hole lifetime as a function of applied magnetic field $B$. At higher fields the direct SLR process caused a decrease of the lifetime with the well-known scaling~$1/B^4$~\cite{Orbach1961,Scott1962,Larson1966,Larson1966a}.

\begin{figure}[t]
\begin{center}
\includegraphics[width=1\linewidth] {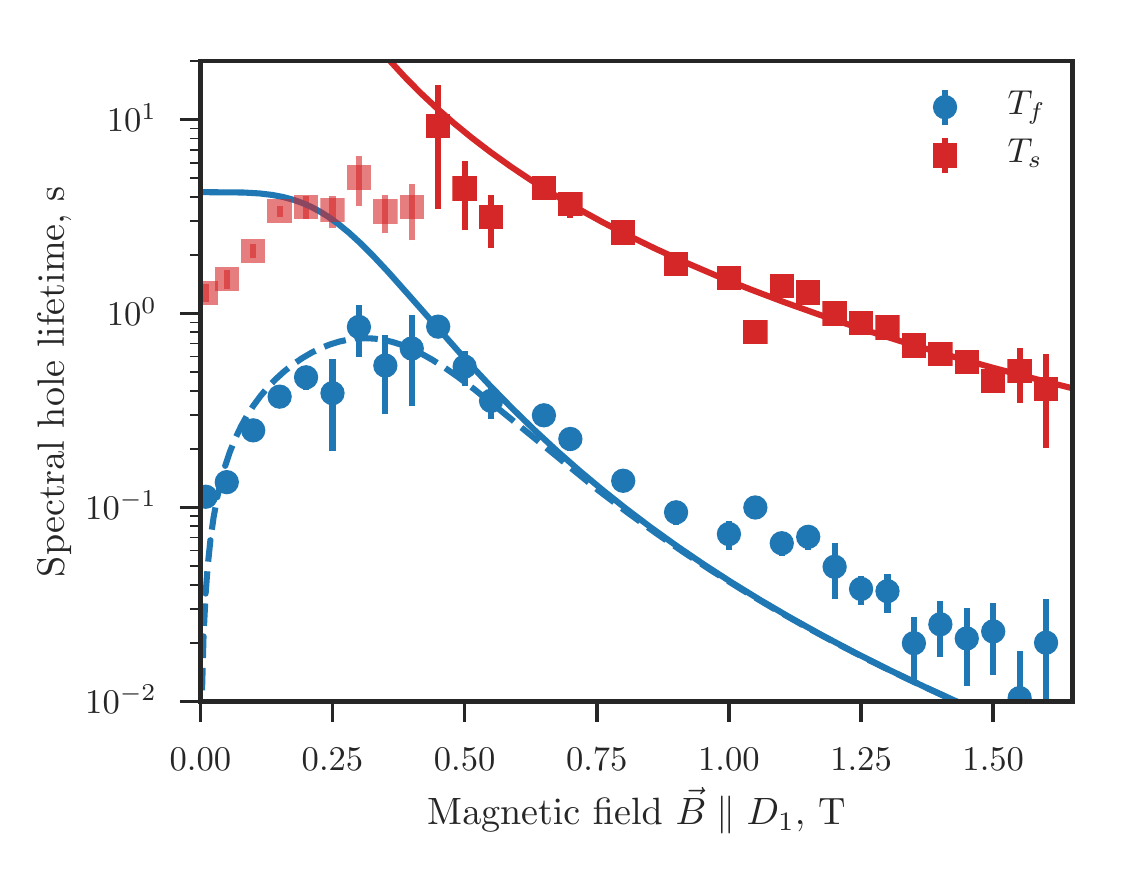}
\caption{The fast $T_f$ (circles) and slow $T_s$ (squares) lifetimes as a function of applied magnetic field along the $D_1$ axis. The blue, solid line shows the SLR model of Ref.~\cite{ZambriniCruzeiro2017a}, using the model parameters fitted in that work. The blue, dashed line shows the complete model of Ref.~\cite{ZambriniCruzeiro2017a}, consisting of the SLR and flip-flop models, where the latter has been scaled down by a factor of 24 as explained in Sec. \ref{sec:relax_model}. The red, solid line is a fitted scaling law $T_s \propto 1/B^{2.5 \pm 0.2}$.}
\label{fig:lifetimes}
\end{center}
\end{figure}

In Figure \ref{fig:lifetimes} the fast and slow lifetimes are shown as a function of the applied field. As seen the lifetimes generally increase at low fields, reach a maximum around 0.4 Tesla, and then start to decrease for higher fields, similarly to the case of ions with $I=0$. In Ref.~\cite{ZambriniCruzeiro2017a} it was found that a SLR model including the direct, Raman and Orbach processes explained well the high-field data for ions with $I=0$. In Figure~\ref{fig:lifetimes} the blue solid line shows the lifetime predicted by this model, using the fitted parameters~\cite{ZambriniCruzeiro2017a} for this particular magnetic field angle ($D_1$ axis). The model agrees rather well with the high-field data of the fast decay component. This is supporting our hypothesis that the fast decay is related to the relaxation process that only changes the electronic spin project ($\Delta m_S = \pm 1$ and $\Delta m_I = 0$), denoted by the rate $R_0$ in Fig.~\ref{fig:model}. The small difference we believe is due to the fact that the $R_0$ should depend slightly on the $m_I$ value, as discussed by Larson and Jeffries~\cite{Larson1966a}, while our rate is an average over all nuclear projections $m_I$. The scaling with magnetic field appears to be somewhat slower than the expected $1/B^4$, possibly also due to the averaging over the $m_I$ states. The experimental measurement uncertainty for these short lifetimes does not allow to be certain, however, that there is a real change in the scaling.

The $B$ scaling of the slower $R_{+}$, $R_{-}$ rates, on the other hand, is expected to be significantly different~\cite{Larson1966a,Abragam1970_page78}. Indeed, as the Zeeman energy increases with the magnetic field, the spin mixing due to the non-secular terms of the hyperfine Hamiltonian will decrease. As discussed by Abragam and Bleaney~\cite{Abragam1970_page78}, this causes the relaxation rates $R_{+}$, $R_{-}$ to decrease, with respect to the $R_0$ relaxation rate, with a factor $B^2$. Hence, qualitatively one would expect a $1/B^2$ scaling of the $R_{+}$, $R_{-}$ rates. In Fig. \ref{fig:lifetimes}, we show a fit of the high-field data of the slow lifetime component to the formula $1/B^a$, which yields an exponent $a = 2.5 \pm 0.2$. As before, the averaging over the $m_I$ states might be responsible for the difference in scaling. Qualitatively, however, the observed scaling supports our hypothesis that the slow decay is related to the SLR rates $R_{+}$, $R_{-}$ involving nuclear spin flips.

Finally, we comment on the lifetimes measured below 0.5 Tesla, where spin flip-flops limit the spectral hole lifetime for ions with $I=0$~\cite{ZambriniCruzeiro2017a}. We first note that for fields down to about 0.15 Tesla, the hyperfine Hamiltonian is still a perturbation to the Zeeman Hamiltonian, in which case one would expect the spin flip-flop process to mainly drive the $R_0$ transition shown in Fig. \ref{fig:model}. This is because the spin flip-flop process is driven by dipole-dipole interactions between the electronic spins, and changes in the nuclear spin projection $m_I$ are due to the hyperfine Hamiltonian, similarly as for the SLR process. Generally we then expect a similar difference in the $R_0$ and $R_+$, $R_-$ rates, respectively. This explains the difference in the measured fast and slow lifetimes also in the low-field region, see Fig. \ref{fig:lifetimes}. In the following we will concentrate on the lifetime of the fast decay.

As compared to the spin flip-flop model that was fitted to the $I=0$ data in Ref.~\cite{ZambriniCruzeiro2017a}, here we expect a much slower flip-flop $R_0$ rate as there are much less spins in the opposite $m_S$ state with the same $m_I$ projection (a reduction of 1/8 due to the $2I+1=8$ nuclear states). In addition our sample is less doped (10 ppm compared to 30 ppm), which reduces the dipole-dipole interaction. In Ref.~\cite{ZambriniCruzeiro2017a} we found approximately a linear dependence of the rate on the Nd$^{3+}$ concentration. So, based on these simple arguments we expect the flip-flop rate to be reduced by factor of 24, qualitatively. This is also supported by the experimental data, as the maximum fast lifetime is 850~ms in this isotopically enriched 10 ppm $^{145}$Nd$^{3+}$:Y$_2$SiO$_5$ crystal, as opposed to 80 ms for the 30 ppm Nd$^{3+}$:Y$_2$SiO$_5$ crystal~\cite{ZambriniCruzeiro2017a}.

In Fig. \ref{fig:lifetimes} we overlay the complete lifetime model of~\cite{ZambriniCruzeiro2017a}, shown as a blue dashed line, including the SLR and flip-flop model, where the flip-flop rate has been scaled down by a factor of 24. The excellent agreement with the scaled-down flip-flop model is rather surprising, given the simplicity of the scaling. Admittedly, however, a more detailed study of the angular dependence of the slow and fast lifetimes would be necessary for a more comprehensive understanding of the flip-flop rate between hyperfine levels. It appears clear, however, that the flip-flop process is a limiting factor also for $^{145}$Nd$^{3+}$, even at concentrations as low as 10~ppm. 

\section{Application to an optical memory}
\label{sec:AFC_storage_application}

In this section we discuss a light storage experiment performed in $^{145}$Nd$^{3+}$:Y$_2$SiO$_5$, based on the atomic frequency comb (AFC) scheme~\cite{Afzelius2009a}. The AFC scheme is based on the creation of a periodic comb structure in the absorption profile, using a periodic SHB procedure. When an optical input pulse is absorbed by the AFC, it will generate an output pulse after a time $1/\Delta$ (the AFC echo), where $\Delta$ is the frequency spacing in the comb. The AFC echo is a delay-line optical memory, which can be used to store quantum states of light~\cite{Saglamyurek2011,Clausen2011,Usmani2012,ZhouHuaLiuEtAl2015}. An on-demand memory can be implemented using a three-level system and optical control pulses~\cite{Afzelius2009a,Afzelius2010,Gundogan2013}, but here we restrict ourselves to the basic AFC echo.

The efficiency of the AFC echo depends on the optical depth $d$ of the comb peaks, the finesse $F$ of the comb and any residual optical depth $d_0$ in between the comb peaks. The efficiency can be calculated using the formula~\cite{Afzelius2009a} $\eta=\tilde{d}^2 \exp(-\tilde{d}) \eta_{deph} \exp(-d_0)$, where $\tilde{d}$ is the average optical depth and $\eta_{deph}$ is an intrinsic dephasing factor due to the comb finesse. The optimal peak shape is squarish~\cite{Bonarota2010}, in which case $\tilde{d}=d/F$ and $\eta_{deph}=\text{sinc}^2(\pi/F)$. For a given optical depth $d$, there is an optimal comb finesse $F=\pi/\arctan (2\pi/d)$. To achieve high efficiencies, the challenge is to make deep spectral holes that preserve the peak optical depth $d$, while reducing $d_0$ such that it does not limit the obtained efficiency significantly. 

\begin{figure}[t]
\begin{center}
\includegraphics[width=1\linewidth] {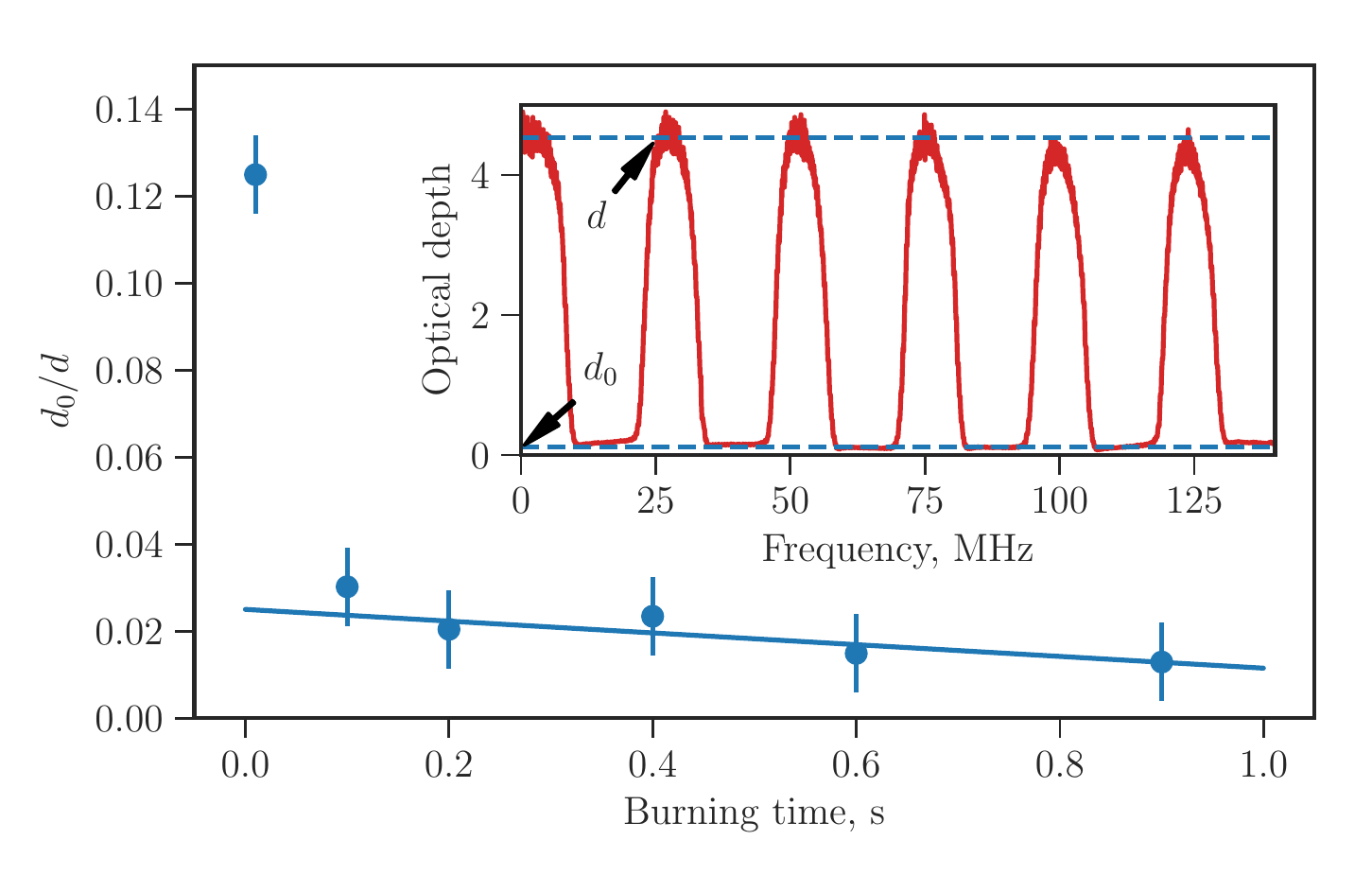}
\caption{Optical pumping efficiency measured by the ratio $d_0/d$, as a function of the burning time $T_{burn}$. For the longest burn time the $d_0/d$ is $1.3 \pm 0.9\%$. The inset shows the AFC structure using a 900 ms burn. The details on the AFC preparation method are described in Ref.~\cite{Clausen2011}.}
\label{fig:afc_spectrum}
\end{center}
\end{figure}

\begin{figure}[b]
\begin{center}
\includegraphics[width=1\linewidth] {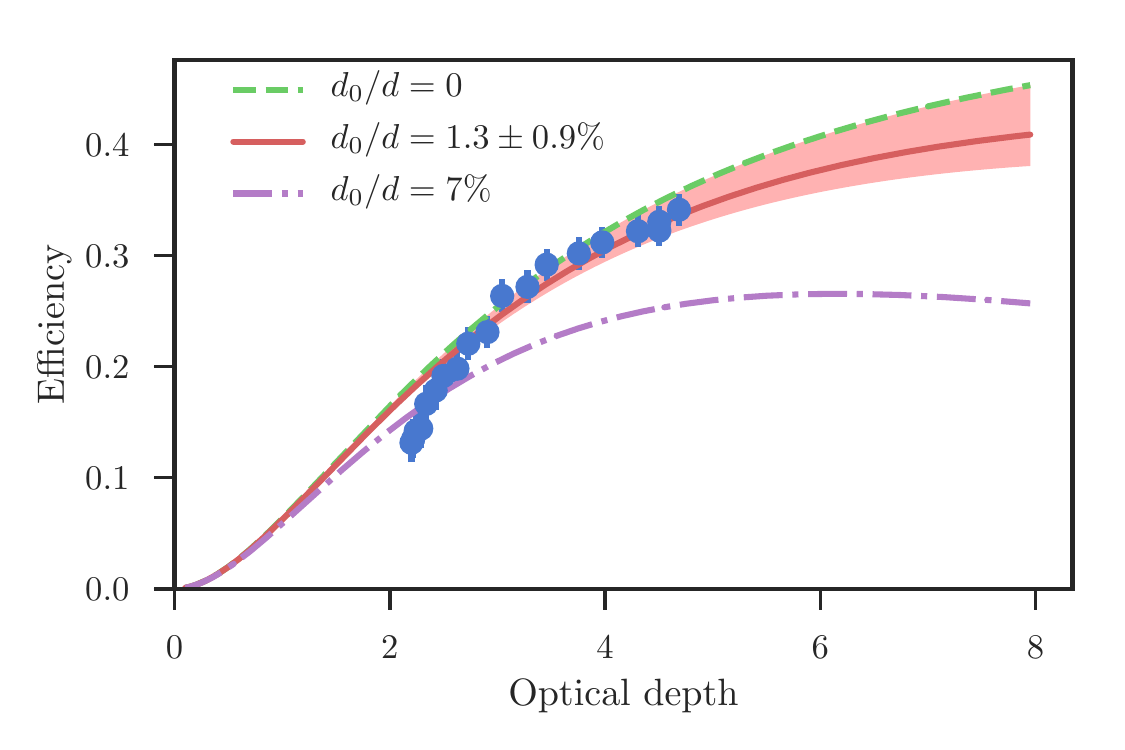}
\caption{The AFC echo efficiency (circles) for different optical depths $d$. The solid (red) line shows the model prediction based on the measured $d_0/d=1.3 \pm 0.9\%$ ratio, with the shaded region showing the uncertainty due to the error estimation in $d_0/d$. The model predictions for the memory reported in~\cite{Clausen2011} ($d_0/d=0.7$) and the perfect memory ($d_0/d=0$) are shown as dot-dashed and dashed lines, respectively.}
\label{fig:afc_efficiency}
\end{center}
\end{figure}

As shown in Fig. \ref{fig:lifetimes}, the maximum lifetimes are reached around a field of 0.4-0.5 T, both for the fast and slow decays. However, we must also consider that the inhomogeneous optical line, as shown in Fig. \ref{fig:abs_spectrum}, splits into two lines for some field due to the ground state splitting, thereby reducing the absorption. For a field along the $D_1$ axis, the ground state $g$-factor is $g=1.47$, such that the split corresponds to the optical inhomogeneous linewidth of 7.7 GHz for a field of 0.37 T. To maximize the absorption we set the field to 0.15 T, where we still reach a spectral hole lifetime of $3.4 \pm 0.2$ s for the slow decay component (see Fig. \ref{fig:lifetimes}). The light also passes through the crystal 4 times, in order to reach a maximum optical depth of $d=4.7$.

In Figure \ref{fig:afc_spectrum} the ratio of the residual to peak optical depth $d_0/d$ is plotted, as a function of the burning time. The comb was created by scanning the laser while modulating the intensity using the double-pass AOM, the same technique used in~\cite{Clausen2011}. As expected, the $d_0/d$ ratio decreases with increasing burn time, while also preserving the peak optical depth, which confirms that the optical pumping efficiency is increasing as the spectral hole decays with a slower rate. For a comb burn time of 900 ms, the ratio is $d_0/d = 1.3 \pm 0.9\%$, significantly lower than the 6-7\% obtained in the reference experiment for Nd$^{3+}$:Y$_2$SiO$_5$ crystals in terms of AFC storage efficiency~\cite{Clausen2011}.

The AFC echo efficiency was then measured as a function of the peak optical depth, using a comb burn time of 1000 ms. The overall optical depth was varied by turning the linear light polarization before the crystal, and using the strong dependence of the optical depth on the polarization~\cite{Clausen2012}. As shown in Fig. \ref{fig:afc_efficiency}, the efficiency reaches a maximum of $33 \pm 1.5\%$ for the maximum optical depth. The measured efficiency is also compared to the theoretical model using the ratio $d_0/d = 1.3 \pm 0.9\%$, showing an excellent agreement. For a longer crystal the model predicts a maximum efficiency of about 43\%. The maximum theoretical efficiency for a ratio of $d_0/d=0.07$ is 26\%, on the other hand, which shows the improvement achieved in the $^{145}$Nd$^{3+}$:Y$_2$SiO$_5$ crystal.

The maximum efficiency is comparable to the {AFC memory} efficiencies reached in praseodymium and europium doped Y$_2$SiO$_5$ crystals without using cavities, which are 35\%~\cite{Amari2010} and 32\%~\cite{Jobez2016}, respectively. Higher efficiencies (50-55\%) have been reached in both materials using cavities~\cite{Jobez2014,Sabooni2013}. An analysis using the equations in Ref.~\cite{Jobez2014} shows that the loss due to the residual absorption would allow cavity-enhanced efficiencies in $^{145}$Nd$^{3+}$:Y$_2$SiO$_5$ in the range of 71-89\%, for $d_0/d$ ratios within the experimental error $1.3 \pm 0.9\%$. The uncertainty in this estimation is quite large due to the difficulty to accurately measure the $d_0/d$ ratio. Also, other losses might also reduce the efficiency~\cite{Jobez2014,Sabooni2013}. The main point is that with spectral hole lifetimes of several seconds, using the hyperfine structure in $^{145}$Nd$^{3+}$, efficient optical pumping is possible and light storage efficiencies can {approach} those reached in non-Kramers ions, where optical pumping is known to be efficient due to long-lived nuclear spin states.

\section{CONCLUSIONS AND OUTLOOK}%

We have presented spectral hole burning decay measurements in an isotopically enriched $^{145}$Nd$^{3+}$:Y$_2$SiO$_5$ crystal. It was observed that the spectral hole decays with two characteristic time scales, a fast and a slow one, which can be explained by a model invoking relaxation paths involving $\Delta m_I=0$ and $\Delta m_I= \pm 1$ transitions, respectively. The magnetic field dependence of the fast and slow decay times was found to follow the same trend as in naturally doped Nd$^{3+}$:Y$_2$SiO$_5$ crystals, but with a reduced contribution from spin flip-flops due to the many hyperfine levels. The maximum lifetime for the slow decay reached about 4 seconds, for a magnetic field of 0.5 Tesla. Furthermore, it was shown that the optical pumping efficiency was improved with respect to naturally doped Nd$^{3+}$:Y$_2$SiO$_5$ crystals, which resulted in AFC echo efficiencies of up to 33\%, similar to those obtained previously in europium and praseodymium crystals. An analysis showed that future experiments could reach efficiencies in the range 71-89\%, by using cavity enhancement. {In order to improve the efficiency beyond such values one would need to further increase the hyperfine lifetime, which could be achieved by optimizing the magnetic field angle and decreasing the dopant concentration \cite{ZambriniCruzeiro2017a}}. Another interesting avenue of research would be to investigate coherent optical-spin manipulation, which would open up the possibility to perform on-demand quantum memory operation in $^{145}$Nd$^{3+}$:Y$_2$SiO$_5$. In this context we emphasize that a nuclear spin coherence times of 9 ms has already been observed in this material~\cite{Wolfowicz2015}. 

\section*{ACKNOWLEDGEMENTS}

We acknowledge useful discussions and some preliminary measurements by F\'{e}lix~Bussi\`{e}res and Thierry~Armici. This work was supported by the European Research Council (ERC-AG MEC), the Swiss programme National Centres of Competence in Research (NCCR) project Quantum Science Technology (QSIT), and the Agence Nationale de la Recherche under grant agreements no. 145-CE26-0037-01 (DISCRYS).

\end{document}